# Skewon field and cosmic wave propagation

Wei-Tou Ni

*Center for Gravitation and Cosmology, Department of Physics,*
*National Tsing Hua University, Hsinchu, Taiwan 30013, Republic of China*
*E-mail:* weitou@gmail.com

**Abstract**

We study the propagation of the Hehl-Obukhov-Rubilar skewon field in weak gravity field/dilute matter or with weak violation of the Einstein Equivalence Principle (EEP), and further classify it into Type I and Type II skewons. From the dispersion relation we show that no dissipation/no amplification condition implies that the additional skewon field must be of Type II. For Type I skewon field, the dissipation/amplification is proportional to the frequency and the CMB spectrum would deviate from Planck spectrum. From the high precision agreement of the CMB spectrum to 2.755 K Planck spectrum, we constrain the Type I cosmic skewon field $|^{(\text{SkI})}\chi^{ijkl}|$ to $\leq$ a few $\times 10^{-35}$. The skewon part of constitutive tensor constructed from asymmetric metric is of Type II, hence is allowed. This study may also find application in macroscopic electrodynamics in the case of laser pumped medium or dissipative medium.

## 1. Introduction

*1.1. Pre-skewon gravitational (spacetime) constitutive tensor density and its constraints*

In the study of how metric arisen from the gravitational interaction of electromagnetism, we started in the 1970's from a constitutive tensor density interaction in analogue to macroscopic electrodynamics. The interaction Lagrangian density $L_\text{I}$ is

$$L_\text{I} = L_\text{I}^{(\text{EM})} + L_\text{I}^{(\text{EM-P})} + L_\text{I}^{(\text{P})} = -(1/(16\pi))\chi^{ijkl} F_{ij} F_{kl} - A_k j^k (-g)^{(1/2)} - \Sigma_I m_I (ds_I)/(dt)\, \delta(\boldsymbol{x}-\boldsymbol{x}_I), \quad (1)$$

with $\chi^{ijkl} = \chi^{klij} = -\chi^{jikl}$ a tensor density of the gravitational fields (e.g., $g_{ij}$, $\varphi$, etc.) or fields to be investigated. Here $F_{ij} \equiv A_{j,i} - A_{i,j}$ is the electromagnetic field strength tensor with $A_i$ the electromagnetic 4-potential and comma denoting partial derivation, $g_{ij}$ is the (symmetric) metric for particles with signature $(+, -, -, -)$, $m_I$ the mass of the $I$th (charged) particle, $s_I$ its 4-line element from the metric $g_{ij}$ with $g$ its determinant, and $j^k(-g)^{(1/2)} \equiv J^k$ is the charge 4-current density. Note that $L_\text{I}^{(\text{EM})}$ and $L_\text{I}^{(\text{EM-P})}$ are metric free ($L_\text{I}^{(\text{EM-P})}$ can be written as $A_k J^k$). In this paper, we use Gaussian units, $c = 1$, and the Einstein summation convention, i.e., summation over repeated indices (indices $i$, $j$, $k$, $l$,… run from 0 to 3) [1, 2]. The gravitational (spacetime) constitutive tensor density $\chi^{ijkl}$ dictates the behaviour of electromagnetism in a gravitational field and has 21 independent components in general. For general relativity or a metric theory (when EEP holds), $\chi^{ijkl}$ is determined completely by the metric $g_{ij}$ and equals $(-g)^{1/2}[(1/2)g^{ik}g^{jl} - (1/2)g^{il}g^{jk}]$; $g^{ik}$ is the inverse of $g_{ik}$ and when replaced by the Minkowski metric $\eta^{ik}$ as in a local inertial frame, we obtain the special relativistic Lagrangian density.

We used two approaches. The first approach used Galileo's Equivalence Principle and derived its consequences [1, 2]. The result is that the constitutive tensor density can be constrained and expressed in metric form with additional pseudoscalar (axion) field $\varphi$:

$$L_I = - (1/(16\pi))(-g)^{1/2}[(1/2)g^{ik}g^{jl}-(1/2)g^{il}g^{kj}+\varphi\ \varepsilon^{ijkl}]F_{ij}F_{kl} - A_k j^k (-g)^{(1/2)} - \Sigma_I m_I (ds_I)/(dt)\delta(\boldsymbol{x}-\boldsymbol{x}_I). \quad (2)$$

The second approach used empirical observations/experiments to constrain the constitutive tensor density interaction [3, 4]. From the no birefringence observations of electromagnetic wave propagation in spacetime, we constructed the light cone and constrained the constitutive tensor to metric form compatible with this light cone plus an additional scalar (dilaton) field and an additional pseudoscalar (axion) field to high precision [3-5]. The theoretical condition for no birefringence (no splitting, no retardation) for electromagnetic wave propagation in all directions is that the constitutive tensor density $\chi^{ijkl}$ can be written in the following form

$$\chi^{ijkl} = (-h)^{1/2}[(1/2)h^{ik}\ h^{jl} - (1/2)h^{il}\ h^{kj}]\psi + \varphi e^{ijkl}, \quad (3)$$

where $h^{ij}$ is a metric constructed from $\chi^{ijkl}$ ($h = \det(h_{ij})$ and $h_{ij}$ the inverse of $h^{ij}$) which generates the light cone for electromagnetic wave propagation [3-6]. $e^{ijkl}$ is the completely anti-symmetric tensor density with $e^{0123} = 1$. We constructed the relation (3) in the weak-violation approximation of the Einstein Equivalence Principle (EEP) in 1981 [3-5]; Haugan and Kauffmann [6] reconstructed the relation (3) in 1995. In 1998, Colladay and Kostelecky [7] used their SME [Standard Model Extension] to study Lorentz violation in Minkowski spacetime. The photon sector of their SME is contained in (1) [8]. In 2004, Lämmerzahl and Hehl [9] laid the cornerstone work of the existence of unique Riemannian light cone from the no birefringence condition. In 2011, Favata and Bergamin [10] finally proved the relation (3) without assuming weak-field approximation (see Dahl [11] also). Polarization measurements of electromagnetic waves from pulsars and cosmologically distant astrophysical sources yield stringent constraints agreeing with (3) down to $2\times10^{-32}$ fractionally (for a review, see [12]).

We further constrained the light cone metric to matter metric up to a scalar factor from Hughes-Drever-type experiments and the dilaton to 1 (constant) from Eötvös-type experiments to high precision [3, 4, 13].

*1.2. Gravitational (spacetime) constitutive tensor density including the skewon field*

Basic formulation of macroscopic electrodynamics is in terms of electromagnetic field tensor $F_{ij}$ and electromagnetic excitation tensor density $H^{ij}$. Maxwell equations expressed in terms of these quantities are

$$H^{ij}{}_{,j} = -4\pi J^i, \tag{4a}$$

$$e^{ijkl} F_{jk,l} = 0 \tag{4b}$$

(See, e. g., Hehl and Obukhov [14]). To complete this set of equations, one needs a constitutive relation between the excitation and the field:

$$H^{ij} = (1/2)\chi^{ijkl} F_{kl}. \tag{5}$$

Since both $H^{ij}$ and $F_{kl}$ are antisymmetric, $\chi^{ijkl}$ must be antisymmetric in $i$ and $j$, and $k$ and $l$. Hence $\chi^{ijkl}$ has 36 independent components. If the Lagrangian formulation as in (1)

$$L_I^{(EM)} = -(1/(8\pi))H^{ij} F_{ij} = -(1/(16\pi))\chi^{ijkl} F_{kl} F_{ij}, \tag{6}$$

is used to derive the equation of motion, only the part of $\chi^{ijkl}$ which is symmetric under the interchange of index pairs $ij$ and $kl$ contributes, and there are effectively only 21 independent functions. If the Lagrangian formulation is not used, the constitutive tensor density $\chi^{ijkl}$ can be asymmetric as in (4)-(5), have 36 independent components, and be decomposed to principal part (P), skewon part (Sk) and axion part (A) as given in [14, 15]:

$$\chi^{ijkl} = {}^{(P)}\chi^{ijkl} + {}^{(Sk)}\chi^{ijkl} + {}^{(A)}\chi^{ijkl}, \qquad (\chi^{ijkl} = -\chi^{jikl} = -\chi^{ijlk}) \tag{7}$$

with

$$^{(P)}\chi^{ijkl} = (1/6)[2(\chi^{ijkl} + \chi^{klij}) - (\chi^{iklj} + \chi^{ljik}) - (\chi^{iljk} + \chi^{jkil})], \tag{8a}$$

$$^{(A)}\chi^{ijkl} = \chi^{[ijkl]} = \varphi\, e^{ijkl}, \tag{8b}$$

$$^{(Sk)}\chi^{ijkl} = (1/2)(\chi^{ijkl} - \chi^{klij}), \tag{8c}$$

The decomposition (7) is unique. The axion interaction has been studied in [16, 17]. Its astrophysical and cosmological constraints are reviewed in [11, 18, 19]. In macroscopic electrodynamics, Hehl, Obukhov, Rivera and Schmid [20] have studied and clarified that the chromium sesquioxide $Cr_2O_3$ is an axionic medium. Their paper has demonstrated that the four-dimensional pseudoscalar $\varphi$ (8b) exists in macroscopic electrodynamics.

## 2. Skewon field and its classification

The Hehl-Obukhov-Rubilar skewon field (8c) can be represented as

$$^{(Sk)}\chi^{ijkl} = e^{ijmk} S_m{}^l - e^{ijml} S_m{}^k, \tag{9}$$

where $S_m{}^n$ is a traceless tensor with $S_m{}^m = 0$ [14]. Explicitly, the correspondence between $^{(Sk)}\chi^{ijkl}$ and $S_m{}^n$ is as follows:

$^{(Sk)}\chi^{1220} = S_3{}^2$; $^{(Sk)}\chi^{1330} = -S_2{}^3$; $^{(Sk)}\chi^{2330} = S_1{}^3$; $^{(Sk)}\chi^{2110} = -S_3{}^1$; $^{(Sk)}\chi^{3110} = S_2{}^1$; $^{(Sk)}\chi^{3220} = -S_1{}^2$;

$^{(Sk)}\chi^{1020} = S_3{}^0$; $^{(Sk)}\chi^{1323} = -S_0{}^3$; $^{(Sk)}\chi^{2030} = S_1{}^0$; $^{(Sk)}\chi^{2131} = -S_0{}^1$; $^{(Sk)}\chi^{3010} = S_2{}^0$; $^{(Sk)}\chi^{3212} = -S_0{}^2$;

$^{(Sk)}\chi^{1320} = -S_0{}^0 - S_2{}^2$; $^{(Sk)}\chi^{1230} = S_0{}^0 + S_3{}^3$; $^{(Sk)}\chi^{2310} = S_0{}^0 + S_1{}^1$; others by symmetry properties.

(10)

We note that $^{(Sk)}\chi^{ijkl}$ has 15 independent degrees of freedom; so is $S_m{}^n$. We can decompose the skewon field into two types – Type I and Type II – by using the Minkowski metric $\eta^{ij}$ (or metric $g^{ij}$) to raise and lower indices of $S_m{}^n$:

I. Type I Skewon field (9 degrees of freedom): $^{(SkI)}\chi^{ijkl}$ with symmetric $^{(SkI)}S_{mn}$,

$$^{(SkI)}S_{mn} = {}^{(SkI)}S_{nm};$$

II. Type II Skewon field (6 degrees of freedom): $^{(SkII)}\chi^{ijkl}$ with anti-symmetric $^{(SkII)}S_{mn}$,

$$^{(SkII)}S_{mn} = - {}^{(SkII)}S_{nm},$$

where

$$S_{mn} \equiv S_m{}^i \eta_{in}; \quad {}^{(SkI)}S_{mn} \equiv (1/2)(S_{mn} + S_{nm}); \quad {}^{(SkII)}S_{mn} \equiv (1/2)(S_{mn} - S_{nm}), \tag{11a}$$

$$S_{mn} = {}^{(SkI)}S_{mn} + {}^{(SkII)}S_{mn}. \tag{11b}$$

This classification is invariant under tensor transformation. General skewon field $^{(Sk)}\chi^{ijkl}$ can be written as the sum of two parts, i.e., $^{(SkI)}\chi^{ijkl} + {}^{(SkII)}\chi^{ijkl}$.

## 3. Constitutive tensor from asymmetric metric

In this section we consider constitutive tensor built from asymmetric metric. Let $q^{ij}$ be the asymmetric metric and resolve it into symmetric part $^{(S)}q^{ij}$ and anti-symmetric part $^{(A)}q^{ij}$:

$$q^{ij} = {}^{(S)}q^{ij} + {}^{(A)}q^{ij}, \text{ with } {}^{(S)}q^{ij} \equiv \tfrac{1}{2}(q^{ij} + q^{ji}) \text{ and } {}^{(A)}q^{ij} \equiv \tfrac{1}{2}(q^{ij} - q^{ji}). \tag{12}$$

The constitutive tensor density can be built from the asymmetric metric $q^{ij}$ as follows:

$$\chi^{ijkl} = \tfrac{1}{2}(-q)^{1/2}(q^{ik}q^{jl} - q^{il}q^{jk}), \tag{13}$$

with $q = \det(^{(S)}q^{ij})$. When $q^{ij}$ is symmetric, this definition reduces to that of the metric theories of gravity. With this definition of $\chi^{ijkl}$, we can resolve it into the principal part plus the axion part $^{(PA)}\chi^{ijkl}$ and skewon part $^{(Sk)}\chi^{ijkl}$ as follows:

$$\chi^{ijkl} = \tfrac{1}{2}(-q)^{1/2}(q^{ik}q^{jl} - q^{il}q^{jk}) = {}^{(PA)}\chi^{ijkl} + {}^{(Sk)}\chi^{ijkl}, \tag{14a}$$

with

$$^{(PA)}\chi^{ijkl} \equiv \tfrac{1}{2}(-q)^{1/2}({}^{(S)}q^{ik}\,{}^{(S)}q^{jl} - {}^{(S)}q^{il}\,{}^{(S)}q^{jk} + {}^{(A)}q^{ik}\,{}^{(A)}q^{jl} - {}^{(A)}q^{il}\,{}^{(A)}q^{jk}), \tag{14b}$$

$$^{(Sk)}\chi^{ijkl} \equiv \tfrac{1}{2}(-q)^{1/2}({}^{(A)}q^{ik}\,{}^{(S)}q^{jl} - {}^{(A)}q^{il}\,{}^{(S)}q^{jk} + {}^{(S)}q^{ik}\,{}^{(A)}q^{jl} - {}^{(S)}q^{il}\,{}^{(A)}q^{jk}). \tag{14c}$$

The axion part of $^{(PA)}\chi^{ijkl}$ only comes from the second order terms of $^{(A)}q^{il}$. This decomposition is also obtained independently by Favaro [21].

Transforming to a coordinate system that $^{(S)}q^{ij}$ takes the form of Minkowski metric $\eta^{ij}$ to the first order in coordinates (hence, $q$ also equals to 1 to first order), we have

$^{(Sk)}\chi^{1220} = S_3{}^2 = - S_{32} = \tfrac{1}{2}\,{}^{(A)}q^{10};\ {}^{(Sk)}\chi^{1330} = - S_2{}^3 = S_{23} = \tfrac{1}{2}\,{}^{(A)}q^{10};\ {}^{(Sk)}\chi^{2330} = S_1{}^3 = - S_{13} = \tfrac{1}{2}\,{}^{(A)}q^{20};$

$^{(Sk)}\chi^{2110} = -S_3{}^1 = S_{31} = \tfrac{1}{2}\,^{(A)}q^{20}$; $^{(Sk)}\chi^{3110} = S_2{}^1 = -S_{21} = \tfrac{1}{2}\,^{(A)}q^{30}$; $^{(Sk)}\chi^{3220} = -S_1{}^2 = S_{12} = \tfrac{1}{2}\,^{(A)}q^{30}$;

$^{(Sk)}\chi^{1020} = S_3{}^0 = S_{30} = \tfrac{1}{2}\,^{(A)}q^{12}$; $^{(Sk)}\chi^{1323} = -S_0{}^3 = S_{03} = -\tfrac{1}{2}\,^{(A)}q^{12}$; $^{(Sk)}\chi^{2030} = S_1{}^0 = S_{10} = \tfrac{1}{2}\,^{(A)}q^{23}$;

$^{(Sk)}\chi^{2131} = -S_0{}^1 = S_{01} = -\tfrac{1}{2}\,^{(A)}q^{23}$; $^{(Sk)}\chi^{3010} = S_2{}^0 = S_{20} = \tfrac{1}{2}\,^{(A)}q^{31}$; $^{(Sk)}\chi^{3212} = -S_0{}^2 = S_{02} = -\tfrac{1}{2}\,^{(A)}q^{31}$;

$^{(Sk)}\chi^{0123}$ and all of its permutation in the upper indices vanish; $S_0{}^0 = S_1{}^1 = S_2{}^2 = S_3{}^3 = 0$. (15)

From (15), we have

$$S_{ij} = \tfrac{1}{4}\, \varepsilon_{ijmk}\,^{(A)}q^{mk};\quad {}^{(A)}q^{mk} = -\varepsilon^{mkij} S_{ij}, \tag{16}$$

where $\varepsilon_{ijmk}$ and $\varepsilon^{mkij}$ are respectively the completely antisymmetric covariant and contravariant tensors with $\varepsilon^{0123} = 1$ and $\varepsilon_{0123} = -1$ in local inertial frame. Thus the skewon field from asymmetric metric is of type II. Its field strength is not limited by the observational consideration of section 5. In $^{(PA)}\chi^{ijkl}$, the axion part only contributes in the second order of $^{(A)}q^{jl}$. Hence in the first order, $^{(PA)}\chi^{ijkl}$ reduces to $^{(P)}\chi^{(1)ijkl}$.

## 4. Wave propagation and the dispersion relation

The sourceless Maxwell equation (4b) is equivalent to the local existence of a 4-potential $A_i$ such that

$$F_{ij} = A_{j,i} - A_{i,j}, \tag{17}$$

with a gauge transformation freedom of adding an arbitrary gradient of a scalar function to $A_i$. The Maxwell equation (4a) in vacuum is

$$(\chi^{ijkl} A_{k,l})_{,j} = 0. \tag{18}$$

Neglecting $\chi^{ijkl}{}_{,m}$ for slowly varying/nearly homogeneous field/medium, or in the eikonal approximation, (18) becomes

$$\chi^{ijkl} A_{k,lj} = 0. \tag{19}$$

In the weak field or dilute medium, we assume

$$\chi^{ijkl} = \chi^{(0)ijkl} + \chi^{(1)ijkl} + \mathrm{O}(2), \tag{20}$$

where O(2) means second order in $\chi^{(1)}$. Since the violation from the Einstein Equivalence Principle would be small and/or if the medium is dilute, in the following we assume that

$$\chi^{(0)ijkl} = (1/2) g^{ik} g^{jl} - (1/2) g^{il} g^{kj}, \tag{21}$$

and $\chi^{(1)ijkl}$ is small compared with $\chi^{(0)ijkl}$. We can then find a locally inertial frame such that $g^{ij}$ becomes the Minkowski metric $\eta^{ij}$ good to the derivative of the metric. To look for wave solutions, we use eikonal approximation and choose z-axis in the wave propagation direction so that the solution takes the following form:

$$A = (A_0, A_1, A_2, A_3)\, \mathrm{e}^{ikz - i\omega t}. \tag{22}$$

We expand the solution as

$$A_i = [A^{(0)}{}_i + A^{(1)}{}_i + \mathrm{O}(2)]\, e^{ikz - i\omega t}. \tag{23}$$

*4.1. Dispersion relation*

Imposing radiation gauge condition in the zeroth order in the weak field/dilute medium/weak EEP violation approximation, we find the zeroth order solution of (23) and the zeroth order dispersion relation satisfying the zeroth order equation $\chi^{(0)ijkl} A^{(0)}{}_{k,lj} = 0$ as follow:

$$A^{(0)} = (0, A^{(0)}{}_1, A^{(0)}{}_2, 0), \quad \omega = k + \mathrm{O}(1). \tag{24}$$

Substituting (23) and (24) into equation (19), we have

$$\chi^{(1)ijkl} A^{(0)}{}_{k,lj} + \chi^{(0)ijkl} A^{(1)}{}_{k,lj} = 0 + \mathrm{O}(2). \tag{25}$$

The $i = 0$ and $i = 3$ components of (24) both give

$$A^{(1)}{}_0 + A^{(1)}{}_3 = 2\,(\chi^{(1)3013} - \chi^{(1)3010})\, A^{(0)}{}_1 + 2\,(\chi^{(1)3023} - \chi^{(1)3020})\, A^{(0)}{}_2 + \mathrm{O}(2). \tag{26}$$

Since this equation does not contain $\omega$ and $k$, it does not contribute to the determination of the dispersion relation. A gauge condition in the O(1) order fixes the values of $A^{(1)}{}_0$ and $A^{(1)}{}_3$.

The $i = 1$ and $i = 2$ components of (25) are

$$(1/2)(\omega^2 - k^2)\, A^{(0)}{}_1 + \chi^{(0)1jkl} A^{(1)}{}_{k,lj} + \chi^{(0)1jkl} A^{(0)}{}_{k,lj} = 0 + \mathrm{O}(2), \tag{27a}$$
$$(-1/2)(\omega^2 - k^2)\, A^{(0)}{}_2 + \chi^{(0)2jkl} A^{(1)}{}_{k,lj} + \chi^{(1)2jkl} A^{(0)}{}_{k,lj} = 0 + \mathrm{O}(2). \tag{27b}$$

These two equations determine the dispersion relation and can be rewritten as

$$[(1/2)(\omega^2 - k^2) - k^2 A_{(1)}]\, A^{(0)}{}_1 - k^2 B_{(1)}\, A^{(0)}{}_2 = \mathrm{O}(2), \tag{28a}$$
$$- k^2 B_{(2)}\, A^{(0)}{}_1 + [(1/2)(\omega^2 - k^2) - k^2 A_{(2)}]\, A^{(0)}{}_2 = \mathrm{O}(2), \tag{28b}$$

where

$$A_{(1)} \equiv \chi^{(1)1010} - (\chi^{(1)1013} + \chi^{(1)1310}) + \chi^{(1)1313}, \tag{29a}$$
$$A_{(2)} \equiv \chi^{(1)2020} - (\chi^{(1)2023} + \chi^{(1)2320}) + \chi^{(1)2323}, \tag{29b}$$
$$B_{(1)} \equiv \chi^{(1)1020} - (\chi^{(1)1023} + \chi^{(1)1320}) + \chi^{(1)1323}, \tag{29c}$$
$$B_{(2)} \equiv \chi^{(1)2010} - (\chi^{(1)2013} + \chi^{(1)2310}) + \chi^{(1)2313}. \tag{29d}$$

We note that $A_{(1)}$ and $A_{(2)}$ contain only the principal part of $\chi$; $B_{(1)}$ and $B_{(2)}$ contain only the *principal and skewon part of $\chi$. The axion part drops out and does not contribute to the dispersion relation in the eikonal approximation.* The principal part $^{(P)}B$ and skewon part $^{(Sk)}B$ of $B_{(1)}$ are as follows:

$$^{(P)}B = (1/2)(B_{(1)} + B_{(2)}); \quad ^{(Sk)}B = (1/2)(B_{(1)} - B_{(2)}). \tag{30}$$

From (30), $B_{(1)}$ and $B_{(2)}$ can be expressed as

$$B_{(1)} = {}^{(P)}B + {}^{(Sk)}B; \quad B_{(2)} = {}^{(P)}B - {}^{(Sk)}B. \tag{31}$$

For equations (28a,b) to have nontrivial solutions of $(A_1^{(0)}, A_2^{(0)})$, we must have the following determinant vanish to first order:

$$\det \begin{bmatrix} (1/2)(\omega^2 - k^2) - k^2 A_{(1)}] & -k^2 B_{(1)} \\ -k^2 B_{(2)} & (1/2)(\omega^2 - k^2) - k^2 A_{(2)} \end{bmatrix}$$
$$= (1/4)(\omega^2 - k^2)^2 - (1/2)(\omega^2 - k^2) k^2 (A_{(1)} + A_{(2)}) + k^4 (A_{(1)} A_{(2)} - B_{(1)} B_{(2)}) = 0 + O(2). \tag{32}$$

The solution of this quadratic equation in $\omega^2$, i.e., the dispersion relation is

$$\omega^2 = k^2 [1 + (A_{(1)} + A_{(2)}) \pm ((A_{(1)} - A_{(2)})^2 + 4B_{(1)} B_{(2)})^{1/2}] + O(2), \tag{33}$$

or

$$\omega = k [1 + 1/2 (A_{(1)} + A_{(2)}) \pm 1/2 ((A_{(1)} - A_{(2)})^2 + 4B_{(1)} B_{(2)})^{1/2}] + O(2). \tag{34}$$

From (33) the group velocity is

$$v_g = \partial\omega/\partial k = 1 + 1/2 (A_{(1)} + A_{(2)}) \pm 1/2 ((A_{(1)} - A_{(2)})^2 + 4B_{(1)} B_{(2)})^{1/2} + O(2). \tag{35}$$

The quantity under the square root sign is

$$\xi \equiv (A_{(1)} - A_{(2)})^2 + 4B_{(1)} B_{(2)} = (A_{(1)} - A_{(2)})^2 + 4({}^{(P)}B)^2 - 4({}^{(Sk)}B)^2. \tag{36}$$

Depending on the sign or vanishing of $\xi$, we have the following three cases of electromagnetic wave propagation:

(i) $\xi > 0$, $(A_{(1)} - A_{(2)})^2 + 4({}^{(P)}B)^2 > 4({}^{(Sk)}B)^2$: There is birefringence of wave propagation;
(ii) $\xi = 0$, $(A_{(1)} - A_{(2)})^2 + 4({}^{(P)}B)^2 = 4({}^{(Sk)}B)^2$: There are no birefringence and no dissipation/amplification in wave propagation;
(iii) $\xi < 0$, $(A_{(1)} - A_{(2)})^2 + 4({}^{(P)}B)^2 < 4({}^{(Sk)}B)^2$: There is no birefringence, but there are both dissipative and amplifying modes in wave propagation.

When the principal part ${}^{(P)}\chi^{ijkl}$ of the constitutive tensor is given by (3), i.e.

$${}^{(P)}\chi^{ijkl} = (-h)^{1/2}[(1/2)h^{ik} h^{jl} - (1/2)h^{il} h^{kj}]\psi, \tag{3a}$$

it is easy to check by substitution that

$$A_{(1)} = A_{(2)} \text{ and } {}^{(P)}B_{(1)} = 0. \tag{3b}$$

In this case, (i) does not happen, (ii) and (iii) become

(ii)' $\xi = 0$, ${}^{(Sk)}B = 0$: There are no birefringence and no dissipation/amplification in wave propagation;
(iii)' $\xi < 0$, ${}^{(Sk)}B \neq 0$: There is no birefringence, but there are both dissipative and amplifying modes in wave propagation.

N.B. For type II skewon field, we have $^{(Sk)}B = 0$ (see next subsection *4.2.*); therefore it is the case (ii)'. In Sec.IV.A.1 of Obukhov and Hehl [22], the pure type II skewon field gives birefringence in the second order of the skewon field strength and there is no birefringence in the first order. This is consistent with our result of no birefringence in the first-order above.

*4.2. The condition of vanishing of $B_{(1)}$ and $B_{(2)}$ for all directions of wave propagation*

The condition of vanishing of $B_{(1)}$ for wave propagation in the *z*-axis direction is

$$B_{(1)} = \chi^{(1)1020} + \chi^{(1)1323} - \chi^{(1)1023} - \chi^{(1)1320} = 0. \tag{37}$$

To look for conditions derivable in combination with those from other directions, we do active Lorentz transformations (rotations/boosts). Active rotation $R_\theta$ in the *y-z* plane with angle $\theta$ is

$$\underline{t} = R_\theta t, \ \underline{x} = R_\theta x, \ \underline{y} = R_\theta y = y \cos \theta + z \sin \theta, \ \underline{z} = R_\theta z = -y \sin \theta + z \cos \theta. \tag{38}$$

Applying active rotation $R_\theta$ (38) to (37), we have

$$\begin{aligned}0 &= \underline{\chi}^{(1)1020} + \underline{\chi}^{(1)1323} - \underline{\chi}^{(1)1023} - \underline{\chi}^{(1)1320} \\ &= \chi^{(1)1020} + \chi^{(1)1323} - \chi^{(1)1023} - \chi^{(1)1320} + \theta \, (\chi^{(1)1030} + \chi^{(1)1220} - \chi^{(1)1223} - \chi^{(1)1330}) + O(\theta^2),\end{aligned} \tag{39}$$

for small value of $\theta$. From (39) and (37), we have

$$\chi^{(1)1030} + \chi^{(1)1220} - \chi^{(1)1223} - \chi^{(1)1330} = 0. \tag{40}$$

Following the same procedure, we apply repeatedly active rotation $R_\theta$ to (40) and the resulting equations together with their linear combinations. After performing cyclic permutation 1→2→3→1 on the upper indices once and twice on some of the resulting equations, we have the following equations (for detailed derivation, see arXiv:1312.3056v1)

$$\chi^{(1)1220} = \chi^{(1)1330}; \ \chi^{(1)2330} = \chi^{(1)2110}; \ \chi^{(1)3110} = \chi^{(1)3220}; \ \chi^{(1)1020} = -\chi^{(1)1323}; \ \chi^{(1)2030} = -\chi^{(1)2131};$$

$$\chi^{(1)3010} = -\chi^{(1)3212}; \ \chi^{(1)1320} = -\chi^{(1)1230}; \ \chi^{(1)2130} = -\chi^{(1)2310}; \ \chi^{(1)3210} = -\chi^{(1)3120}; \ \chi^{(1)1023} = -\chi^{(1)1320};$$

$$\chi^{(1)2031} = -\chi^{(1)2130}; \ \chi^{(1)3012} = -\chi^{(1)3210}. \tag{41a-l}$$

From (41g-l), $\chi^{(1)0123}$ is completely anti-symmetric under any permutation of (0123). Among (41g-i) only 2 are independent; among (41j-l) also only 2 are independent. For $^{(PA)}\chi^{ijkl}$, (41g-l) give 2 independent conditions. For $^{(Sk)}\chi^{ijkl}$, (41g-l) give 3 independent conditions and $\chi^{(1)0123}$ must vanish.

The derivation of formulas in this subsection from (37) to (41l) is independent of whether $\chi^{ijkl}$ is principal, axionic or skewonic. Hence, $^{(P)}$(41a-l) hold for $^{(P)}\chi^{ijkl}$ with $^{(P)}B_{(1)} = 0$, $^{(A)}$(41a-l) hold for $^{(A)}\chi^{ijkl}$ with $^{(A)}B_{(1)} = 0$, and $^{(Sk)}$(41a-l) holds for $^{(Sk)}\chi^{ijkl}$ with $^{(Sk)}B_{(1)} = 0$. Here $^{(P)}$(41a-l) means (41a-l) with $\chi$ substituted by $^{(P)}\chi$, $^{(A)}$(41a-l) means (41a-l) with $\chi$ substituted by $^{(A)}\chi$, and

$^{(Sk)}$(41a-l) means (41a-l) with $\chi$ substituted by $^{(Sk)}\chi$; similarly for $^{(P)}B_{(1)}$, $^{(A)}B_{(1)}$ and $^{(Sk)}B_{(1)}$. For $B_{(1)} = B_{(2)} = 0$ in all directions, we have $^{(P)}B_{(1)} = {}^{(Sk)}B_{(1)} = 0$ in all directions, and hence, both $^{(P)}$(41a-l) and $^{(Sk)}$(41a-l) are valid.

The last 3 equalities of (10) written in first order form are

$$^{(Sk)}\chi^{(1)1320} = -S^{(1)0}{}_0 - S^{(1)2}{}_2;\ {}^{(Sk)}\chi^{(1)1230} = S^{(1)0}{}_0 + S^{(1)3}{}_3;\ {}^{(Sk)}\chi^{(1)2310} = S^{(1)0}{}_0 + S^{(1)1}{}_1. \tag{42}$$

From $^{(Sk)}$(41g-41l), we must have $^{(Sk)}\chi^{(1)1320} = {}^{(Sk)}\chi^{(1)1230} = {}^{(Sk)}\chi^{(1)2310} = 0$. From (42) and Tr $S_n{}^m = 0$, then all $S^{(1)0}{}_0$, $S^{(1)1}{}_1$, $S^{(1)2}{}_2$ and $S^{(1)3}{}_3$ must vanish.

The first 12 equalities of (10) written in first order form together with $^{(Sk)}$(41a-41f) give

$$S^{(1)2}{}_3 = -S^{(1)3}{}_2;\ S^{(1)3}{}_1 = -S^{(1)1}{}_3;\ S^{(1)1}{}_2 = -S^{(1)2}{}_1;\ S^{(1)0}{}_3 = S^{(1)3}{}_0;\ S^{(1)0}{}_1 = S^{(1)1}{}_0;\ S^{(1)0}{}_2 = S^{(1)2}{}_0. \tag{43}$$

Using the Lorentz metric to raise/lower the indices, we have

$$S^{(1)mn} = -S^{(1)nm},\ S^{(1)}{}_{mn} = -S^{(1)}{}_{nm}. \tag{44}$$

Thus, when $^{(Sk)}$(41a-41l) (9 independent conditions) are satisfied, the skewon degrees of freedom are reduced to 6 (15 – 9) and only Type II skewon field remains.

Under Lorentz transformation, the symmetric part and the anti-symmetric part of $S^{mn}$ transform separately. Hence, with the conditions $^{(Sk)}B = 0$ for all directions of wave propagation, the skewon field is constrained to Type II. The reverse is also true: Since $^{(SkII)}S_{nm}$ is a tensor, when it satisfy $^{(Sk)}B = 0$ for the $z$-axis of wave propagation, they satisfy $^{(Sk)}B = 0$ for *all directions* of wave propagation. Hence we have the lemma:

*Lemma*: The following three statements are equivalent

    (i) $^{(Sk)}B = 0$ for all directions,

    (ii) $^{(Sk)}$(41a-41l) hold,

    (iii) $^{(Sk)}S_{mn}$ as defined by (9) can be written as $^{(Sk)}S_{mn} = {}^{(SkII)}S_{mn}$ with $^{(SkII)}S_{nm} = -{}^{(SkII)}S_{mn}$.

Proof: (i) → (ii) has been demonstrated in the derivation of $^{(Sk)}$(41a-41l).

    (ii) ↔ (iii) has also been demonstrated in the derivation of (42)-(44) and its reversibility.

    (iii) → (i). $^{(SkII)}S_{ij}$ is a Lorentz tensor. If its anti-symmetric property holds in one Lorentz frame, it holds in any Lorentz frame. Hence, in any new Lorentz frame with the propagation in the $\underline{z}$-direction, $^{(Sk)}$(41a-41l) hold and we have $^{(Sk)}\underline{B} = 0$ for propagation in the $\underline{z}$-direction. Since $\underline{z}$-direction can be arbitrary, we have $^{(Sk)}\underline{B} = 0$ for all directions.

### 4.3. *The condition of $^{(Sk)}B_{(1)} = {}^{(P)}B_{(1)} = 0$ and $A_{(1)} = A_{(2)}$ for all directions of wave propagation*

With the condition $^{(Sk)}B_{(1)} = {}^{(P)}B_{(1)} = 0$ and $A_{(1)} = A_{(2)}$ for all directions of wave propagation, there is no birefringence for all directions of wave propagation and we are ready to generalize equation (3) in the case $^{(Sk)}\chi$ is included. From subsection 4.2, we have equations (41a-l) holds from the validity of $^{(Sk)}B_{(1)} = {}^{(P)}B_{(1)} = 0$ (i.e., $B_{(1)} = 0$) for all directions of wave propagation. From $A_{(1)} = A_{(2)}$ and the definition (29a, b), we have

$$\chi^{(1)1010} - (\chi^{(1)1013} + \chi^{(1)1310}) + \chi^{(1)1313} = \chi^{(1)2020} - (\chi^{(1)2023} + \chi^{(1)2320}) + \chi^{(1)2323}. \tag{45}$$

From (47c) for the principal part, the terms in the parentheses on the two sides of the above equation cancel out and we have

$$\chi^{(1)1010} + \chi^{(1)1313} = \chi^{(1)2020} + \chi^{(1)2323}. \tag{46a}$$

Applying active rotation $R_{\pi/2}$ around in the $y$-$z$ plane to (46a), we obtain

$$\chi^{(1)1010} + \chi^{(1)1212} = \chi^{(1)3030} + \chi^{(1)\,3232}. \tag{46b}$$

Define

$$h^{(1)10} \equiv h^{(1)01} \equiv -2\,{}^{(P)}\chi^{(1)1220};\ h^{(1)20} \equiv h^{(1)02} \equiv -2\,{}^{(P)}\chi^{(1)2330};\ h^{(1)30} \equiv h^{(1)03} \equiv -2\,{}^{(P)}\chi^{(1)3110};$$

$$h^{(1)12} \equiv h^{(1)21} \equiv -2\,{}^{(P)}\chi^{(1)1020};\ h^{(1)23} \equiv h^{(1)32} \equiv -2\,{}^{(P)}\chi^{(1)2030};\ h^{(1)31} \equiv h^{(1)13} \equiv -2\,{}^{(P)}\chi^{(1)3010};$$

$$h^{(1)11} \equiv 2\,{}^{(P)}\chi^{(1)2020} + 2\,{}^{(P)}\chi^{(1)2121} - h^{(1)00};\ h^{(1)22} \equiv 2\,{}^{(P)}\chi^{(1)3030} + 2\,{}^{(P)}\chi^{(1)3232} - h^{(1)00};$$

$$h^{(1)33} \equiv 2\,{}^{(P)}\chi^{(1)1010} + 2\,{}^{(P)}\chi^{(1)1313} - h^{(1)00}, \tag{47a}$$

$$\psi \equiv 1 + 2\,{}^{(P)}\chi^{(1)1212} + (1/2)\,\eta_{00}\,(h^{(1)00} - h^{(1)11} - h^{(1)22} - h^{(1)33}) - h^{(1)11} - h^{(1)22}, \tag{47b}$$

$$\varphi \equiv \chi^{(1)0123} \equiv \chi^{(1)[0123]}. \tag{47c}$$

Note that in these definitions, $h^{(1)00}$ is not defined and is free. Define the anti-symmetric metric $p^{ij}$ as follow:

$$p^{10} \equiv -p^{01} \equiv 2\,{}^{(SkII)}\chi^{(1)1220};\ p^{20} \equiv -p^{02} \equiv 2\,{}^{(SkII)}\chi^{(1)2330};\ p^{30} \equiv -p^{03} \equiv 2\,{}^{(SkII)}\chi^{(1)3110};$$

$$p^{12} \equiv -p^{21} \equiv 2\,{}^{(SkII)}\chi^{(1)1020};\ p^{23} \equiv -p^{32} \equiv 2\,{}^{(SkII)}\chi^{(1)2030};\ p^{31} \equiv -p^{13} \equiv 2\,{}^{(SkII)}\chi^{(1)3010};$$

$$p^{00} \equiv p^{11} \equiv p^{22} \equiv p^{33} \equiv 0. \tag{47d}$$

It is straightforward to show now that when (41a-l) and (46a-b) are satisfied, then $\chi$ can be written to first-order in terms of the fields $h^{(1)ij}$, $\psi$, $\varphi$, and $p^{ij}$ with $h^{ij} \equiv \eta^{ij} + h^{(1)ij}$ and $h \equiv \det(h_{ij})$ in the following form:

$$\chi^{ijkl} = {}^{(P)}\chi^{(1)ijkl} + {}^{(A)}\chi^{(1)ijkl} + {}^{(SkII)}\chi^{(1)ijkl}$$
$$= \tfrac{1}{2}\,(-h)^{1/2}[h^{ik}h^{jl} - h^{il}h^{kj}]\psi + \varphi e^{ijkl} + \tfrac{1}{2}\,(-\eta)^{1/2}\,(p^{ik}\eta^{jl} - p^{il}\eta^{jk} + \eta^{ik}p^{jl} - \eta^{il}p^{jk}), \tag{48}$$

with

$${}^{(P)}\chi^{(1)ijkl} = \tfrac{1}{2}\,(-h)^{1/2}[h^{ik}h^{jl} - h^{il}h^{kj}]\psi, \tag{49a}$$

$${}^{(A)}\chi^{(1)ijkl} = \varphi e^{ijkl}, \tag{49b}$$

$${}^{(SkII)}\chi^{(1)ijkl} = \tfrac{1}{2}\,(-\eta)^{1/2}\,(p^{ik}\eta^{jl} - p^{il}\eta^{jk} + \eta^{ik}p^{jl} - \eta^{il}p^{jk}). \tag{49c}$$

Now we are ready to derive the following theorem:

*Theorem*: With $^{(Sk)}B = 0$ for all directions, the following three statements are equivalent to first order in the field:

    (i) $A_{(1)} = A_{(2)}$ and $^{(P)}B = 0$ for all directions,

    (ii) $^{(P)}$(41a-l) and (46a-b) hold,

    (iii) $\chi^{ijkl}$ can be expressed as (48) with (49a-c).

Proof: (i) → (ii) has been demonstrated in the derivation of $^{(P)}$(41a-41l) and (46a-b).

  (ii) → (iii) has also been demonstrated in the derivation of (48) above.

  (iii) → (i): (48) is a Lorentz tensor density equation. If it holds in one Lorentz frame, it holds in any other frame. From this we readily check that $A_{(1)} = A_{(2)}$ and $^{(P)}B = 0$ in any new frame with the wave propagation in the $z$-direction.

As a corollary, *for symmetric $\chi^{ijkl}$* as in (1), we have $^{(Sk)}\chi^{ijkl} = 0$ and $^{(Sk)}B = 0$ automatically satisfied, the following three statements are equivalent to first order:

    (i) $A_{(1)} = A_{(2)}$ and $B = 0$ for all directions, i.e. no birefringence for all directions,

    (ii) (41a-l) and (46a-b) hold,

    (iii) $\chi^{ijkl}$ can be expressed as (3).

With the corollary, we recovered the results of our former work [3-5]. We note that previously we used the symbol $H^{ik}$ instead of $h^{ik}$, here because $H^{ik}$ is already used for excitation, we changed the notation.

## 5. No amplification/dissipation constraint to ultrahigh precision from CMB observations

In this section we look into the observations/experiments to constrain the skewon field contribution to spacetime constitutive tensor density. If $\xi$ as defined in (36) is less than zero, i.e. $(A_{(1)} - A_{(2)})^2 + 4(^{(P)}B)^2 < 4(^{(Sk)}B)^2$, the dispersion relation (34) is

$$\omega = k\,[1 + \tfrac{1}{2}\,(A_{(1)} + A_{(2)}) \pm \tfrac{1}{2}\,(-\xi)^{1/2}i] + O(2). \tag{50}$$

The exponential factor in the wave solution (23) is of the form

$$\exp(ikz - i\omega t) \sim \exp[ikz - ik\,(1 + 1/2\,(A_{(1)} + A_{(2)}))\,t]\,\exp(\pm\tfrac{1}{2}\,(-\xi)^{1/2}kt). \tag{51}$$

There are both dissipative and amplifying wave propagation modes. In the small $\xi$ limit, the amplification/attenuation factor $\exp(\pm\tfrac{1}{2}(-\xi)^{1/2}kt)$ equals $[1 \pm \tfrac{1}{2}(-\xi)^{1/2}kt]$ to a very good approximation. Since this factor depends on the wave number/frequency, it will distort the source spectrum in propagation.

The spectrum of the cosmic microwave background (CMB) is well understood to be Planck blackbody spectrum. It is measured to agree with the ideal Planck spectrum at temperature $2.7255 \pm 0.0006$ K [23]. The measured shape of the CMB spectra does not deviate from Planck spectrum within its experimental accuracy. The agreement for the overall shape with a fit to Planck plus a linear factor $[1 \pm \frac{1}{2}(-\xi)^{1/2}kt]$ is to agree with Planck to better than $10^{-4}$. Planck Surveyor has nine bands of detection from 30 to 857 GHz [24]. For weak propagation deviation, the amplitude of the wave is increased or decreased linearly as $\frac{1}{2}(-\xi)^{1/2}kt$ depending on frequency. For cosmic propagation, the CMB amplitude change due to redshift (or blue shift) is universal. The frequency (wave number) change is proportional to $(1 + z(t))$ with $z(t)$ the redshift factor at time $t$ of propagation. We need to replace $kt$ in the $[1\pm \frac{1}{2}(-\xi)^{1/2}kt]$ factor by the integral

$$\int k(t)\, dt = \int k(t_0)\, (1+ z(t))\, dt \equiv (1+ <z(t)>)\, k(t_0)\, (t_0 - t_1), \tag{52}$$

with $<z(t)>$ the average of $z(t)$ during propagation defined by the last equality of (528), $t_0$ the present time (the age of our universe), and $t_1$ the time at the photon decoupling epoch. According to *Planck* 2013 results [24], the age of our universe $t_0$ is 13.8 Gyr, the decoupling time $t_1$ is 0.00038 Gyr, hence $(t_0 - t_1)$ is ~13.8 Gyr, and $z(t_1)$ is 1090. Using *Planck* ΛCDM concordance model, the factor $(1+ <z(t)>)$ is estimated to be about 3 and the value $(1+ <z(t)>)(t_0 - t_1)$ is more than 40 Gyr. The factor $(1+ <z(t)>)$ multiply by $(t_0 - t_1)$ is the angular diameter distance $D_A$ at which we are observing the CMB and is equal to the comoving size of the sound horizon at the time of last-scattering, $r_s(z(t_1))$, divided by the observed angular size $\theta_* = r_s/D_A$ from seven acoustic peaks in the CMB anisotropy spectrum. From Planck results, $r_s = 144.75 \pm 0.66$ Mpc and $\theta_* = (1.04148 \pm 0.00066) \times 10^{-2}$. Hence, we have $D_A = r_s/\theta_* = 13898 \pm 64$ Mpc $= 45.328 \pm 0.21$ Gyr. This is consistent with our integral estimation.

For the highest frequency band $\omega$ is $2\pi \times 857$ GHz. The amplification/dissipation in fraction is

$$\frac{1}{2}(-\xi)^{1/2}k \times 45.328 \text{ Gyr} = 3.8 \times 10^{30}(-\xi)^{1/2}. \tag{53}$$

For the lowest frequency band $\omega$ is $2\pi \times 30$ GHz; the effect is about ±3.5 % of (53). From CMB observations that the spectrum is less than $10^{-4}$ deviation, we have

$$(-\xi)^{1/2} < 2.6 \times 10^{-35}. \tag{54}$$

When the spacetime constitutive tensor is constructed from metric, dilaton and axion plus skewon, the principal part $^{(P)}\chi^{ijkl}$ of the constitutive tensor is given by (3) (i.e., (3a)) and it is easy to check by substitution that $A_{(1)} = A_{(2)}$ and $^{(P)}B_{(1)} = 0$. There are two cases, (ii)' $^{(Sk)}B = 0$ and (iii)' $^{(Sk)}B \neq 0$ as mentioned at the end of subsection *4.1*. For case (ii)' $\xi = 0$, there are no birefringence and no dissipation/amplification in wave propagation; by the *Theorem* in subsection *4.3.*, the skewon part must be of Type II. For case (iii)' $\xi < 0$, $^{(Sk)}B \neq 0$, there are both dissipative and amplifying modes in wave propagation and we can apply the (54) from the CMB observations to constrain the skewon part of the constitutive tensor as follows

$$\frac{1}{2}(-\xi)^{1/2} = |^{(Sk)}B| = \frac{1}{2}|(B_{(1)} - B_{(2)})| = |^{(Sk)}\chi^{(1)1020} + {}^{(Sk)}\chi^{(1)1323} - {}^{(Sk)}\chi^{(1)1023} - {}^{(Sk)}\chi^{(1)1320}| < 1.3 \times 10^{-35}, \tag{55}$$

for propagation in the *z*-direction. Since the CMB observation is omnidirectional, we have the above constraint for many directions. From a few superpositions, we obtain the *Lemma* in subsection *4.2.*, hence the constraints (41a-l) hold to ~ a few × $20^{-35}$ and the spacetime skewon field is Type II with type I skewon field constrained to ~ a few × $20^{-35}$ cosmologically. Thus, *the significant skewon field must be of Type II with six degrees of freedom.*

## 6. Discussion and outlook

*The main conclusion of this paper is that Type I skewon field is stringently constrained by the CMB spectrum while Type II skewon field may be allowed as part of spaetime structure.* This study may also find application in macroscopic electrodynamics in the case of laser pumped medium or dissipative medium. In the following, we highlight various points and discuss various issues to be investigated further.

(i) We have studied the Hehl-Obukhov-Rubilar skewon field as a possible component of spacetime constitutive tensor. First, we classify the skewon field $S_m{}^n$ into Type I and Type II with the help of a spacetime metric. Type I is symmetric in $S_{mn}$; type II is antisymmetric in $S_{mn}$.

(ii) In the spacetime propagation of electromagnetic field, the skewon field, in general, induces amplification/dissipation. From the precise agreement of the CMB spectrum with the Planck blackbody spectrum, we put an upper limit about a few × $20^{-35}$ on the strength of cosmic skewon field of type I. Direct fitting of CMB data with a multiplicative factor $[1 \pm (-\xi)^{1/2}kt]$ to determine the $(-\xi)^{1/2}$ may give better precision. Looking for propagation of higher frequency electromagnetic radiation of well-understood astrophysical and cosmological sources may further improve the precision.

(iii) Type II skewon field are not constrained in the first order. Further general studies on the properties of skewon field, especially on the electromagnetic propagation in finite-amplitude skewon field, along the line of references [25-27] are critical for considering the combined effects of $A_{(1)}$, $A_{(2)}$, $^{(P)}B$ and $^{(Sk)}B$ on birefringence and amplification/dissipation. This is especially important for applications in macroscopic electrodynamics with pumped/dissipation media.

(iv) The skewon field constructed from asymmetric metric is of type II and passes the experimental test of section 5.

(vi) In the case of no skewon field, from the no birefringence condition in all directions, a light cone metric can be constructed (plus a dilaton field and an axion field) from previous investigations [3-6, 9-11]. In section 4, we generalize this theorem to the case of $^{(Sk)}B = 0$ for all directions, and found that from the 'no birefringence condition' (in all directions), we can construct an asymmetric metric plus a dilation field and an axion field (in the weak field/weak EEP violation approximation). It would be interesting to work out what are the implications from the 'no birefringence condition' (in all directions) and the 'no amplification/dissipation' condition (in all directions) for general case (i.e., the case of $^{(Sk)}B \neq 0$ and non-weak field ).

(vii) In view of the existence of axionic material [20], searching for skewonic material in macroscopic electrodynamics would also be interesting. Mutual stimulations are critical. In physics community, the condensed-matter Higgs particles are actively under study [28]; the Majorana fermions in superconductors are actively searched for also [29].

(viii) Since gravity is universal, in previous papers [30, 31], we have explored the foundations of classical electrodynamics using $\chi$-$g$ formalism. In this paper, we have extended the approach to include skewon field. CMB observations constrained the Type I skewon field ultra-stringently. However, we found Type II skewon field may be allowed and demands more studies.

**Acknowledgements**

We would like to thank A. Favaro, F. W. Hehl and Yu. N. Obukhov for helpful discussions on the skewon field and on various previous works. We would also like to thank the National Science Council (Grants No. NSC101-2112-M-007-007 and No. NSC102-2112-M-007-019) and the National Center for Theoretical Sciences (NCTS) for supporting this work in part.

**References**

[1] W.-T. Ni, *Phys. Rev. Lett.* 38 (1977) 301–304.

[2] W.-T. Ni, *Bull. Am. Phys. Soc.* 19 (1974) 655.

[3] W.-T. Ni, Equivalence Principles, Their Empirical Foundations, and the Role of Precision Experiments to Test Them, in *Proceedings of the 1983 International School and Symposium on Precision Measurement and Gravity Experiment*, Taipei, Republic of China, January 24-February 2, 1983, ed. by W.-T. Ni (Published by National Tsing Hua University, Hsinchu, Taiwan, Republic of China, 1983) pp. 491-517.

[4] W.-T. Ni, Equivalence Principles and Precision Experiments, in *Precision Measurement and Fundamental Constants II*, ed. by B. N. Taylor and W. D. Phillips, Natl. Bur. Stand. (U S) Spec. Publ. 617 (1984) pp. 647-651 [http://astrod.wikispaces.com/].

[5] W.-T. Ni, Timing Observations of the Pulsar Propagations in the Galactic Gravitational Field as Precision Tests of the Einstein Equivalence Principle, in *Proceedings of the Second Asian-Pacific Regional Meeting of the International Astronomical Union on Astronomy, Bandung, Indonesia – 24 to29 August 1981*, ed. by B. Hidayat and M. W. Feast (Published by Tira Pustaka, Jakarta, Indonesia, 1984) pp. 441-448.

[6] M. Haugan and T. Kauffmann, Phys. Rev. D 52 (1995) 3168.

[7] D. Colladay and V. A. Kostelecký, Phys. Rev. D 58 (1998) 116002.

[8] The photon sector of the SME Lagrangian is given by $\mathcal{L}_{\text{photon}}^{\text{total}} = -(1/4) F_{\mu\nu} F^{\mu\nu} - (1/4) (k_F)_{\kappa\lambda\mu\nu} F^{\kappa\lambda} F^{\mu\nu} + (1/2) (k_{AF})^\kappa \varepsilon_{\kappa\lambda\mu\nu} A^\lambda F^{\mu\nu}$ (equation (31) of [7]). The CPT-even part ($-(1/4) (k_F)_{\kappa\lambda\mu\nu} F^{\kappa\lambda} F^{\mu\nu}$) has

constant components $(k_F)_{\kappa\lambda\mu\nu}$ which correspond one-to-one to our $\chi$'s when specialized to constant values minus the special relativistic $\chi$ with the constant axion piece dropped, i.e. $(k_F)^{\kappa\lambda\mu\nu} = \chi^{\kappa\lambda\mu\nu} - (1/2)(\eta^{\kappa\mu}\eta^{\lambda\nu} - \eta^{\kappa\nu}\eta^{\lambda\mu})$. The CPT-odd part $(k_{AF})^\kappa$ also has constant components which correspond to the derivatives of axion $\varphi,^\kappa$ when specialized to constant values.

[9] C. Lämmerzahl and F. W. Hehl, *Phys. Rev. D* 70 (2004) 105022.

[10] A. Favaro and L. Bergamin, Annalen der Physik 523 (2011) 383-401.

[11] M. F. Dahl, Journal of Physics A: Mathematical and Theoretical 45 (2012) 405203.

[12] W.-T. Ni, *Reports on Progress in Physics* 73 (2010) 056901.

[13] W.-T. Ni, Implications of Hughes-Drever Experiments, in *Proceedings of the 1983 International School and Symposium on Precision Measurement and Gravity Experiment*, Taipei, Republic of China, January 24-February 2, 1983, ed. by W.-T. Ni (Published by National Tsing Hua University, Hsinchu, Taiwan, Republic of China, 1983) pp. 519-529 [http://astrod.wikispaces.com/].

[14] F. W. Hehl and Yu. N. Obukhov, *Foundations of Classical Electrodynamics: Charge, Flux, and Metric* (Birkhäuser: Boston, MA, 2003).

[15] F. W. Hehl, Yu. N. Obukhov, G. F. Rubilar, On a possible new type of a T-odd skewon field linked to electromagnetism, in: A. Macias, F. Uribe, E. Diaz (Eds.), Developments in Mathematical and Experimental Physics, Volume A: Cosmology and Gravitation (Kluwer Academic/Plenum, New York, 2002) pp. 241-256 [gr-qc/0203096].

[16] W.-T. Ni, A Nonmetric Theory of Gravity, preprint, Montana State University (1973) [http://astrod.wikispaces.com/].

[17] W.-T. Ni, Spin, Torsion and Polarized Test-Body Experiments, in *Proceedings of the 1983 International School and Symposium on Precision Measurement and Gravity Experiment*, Taipei, Republic of China, January 24-February 2, 1983, ed. by W.-T. Ni (Published by National Tsing Hua University, Hsinchu, Taiwan, Republic of China) pp. 531-540 [http://astrod.wikispaces.com/].

[18] S. di Serego Alighieri, Cosmological Birefringence: an Astrophysical test of Fundamental Physics, Proceeding of Symposium I of JENAM 2010 – Joint European and National Astronomy Meeting: From Varying Couplings to Fundamental Physics, Lisbon, 6-10 Sept. 2010, Editors C. Martins and P. Molaro (Springer-Verlag, Berlin, 2011) p.139 [arXiv:1011.4865].

[19] W.-T. Ni, *Prog. Theor. Phys. Suppl.* **172** (2008) 49 [arXiv:0712.4082].

[20] F. W. Hehl, Yu. N. Obukhov, J.-P. Rivera and H. Schmid, Phys. Rev. A 77 (2008) 022106.

[21] A. Favaro, Recent advances in classical electromagnetic theory, PhD thesis, Imperial College London, 2012. We thank him for sending a copy this reference immediately after our paper was submitted to arXiv [arXiv:1312.3056v1].

[22] Yu. N. Obukhov and F. W. Hehl, Phys. Rev. D 70 (2004) 125015.

[23] D. J. Fixsen, Astrophys. J. 707 (2009) 916.

[24] P. A. R. Ade *et al*. (Planck Collaboration), Planck 2013 results. XVI. Cosmological parameters, arXiv:1303.5076v1 [astro-ph.CO].

[25] F. W. Hehl, Yu. N. Obukhov, G. F. Rubilar and M. Blagojevic, Phys. Lett. A 347 (2005) 14.

[26] C. Lämmerzahl, A. Macías and H. Müller, Phys. Rev. D 71 (2005) 025007.

[27] Y. Itin, Phys. Rev. D 88 (2013) 107502.

[28] E. S. Reich, Nature 495 (2013) 422.

[29] C. W. J. Beenakker, Annu. Rev. Condens. Matter Phys. 4 (2013) 113-136.

[30] W.-T. Ni, Foundations of Electromagnetism, Equivalence Principles and Cosmic Interactions, Chaper 3 in *Trends in Electromagnetism - From Fundamentals to Applications*, pp. 45-68 (March, 2012), Victor Barsan and Radu P. Lungu (Ed.), ISBN: 978-953-51-0267-0, InTech (open access) (2012) [arXiv:1109.5501], Available from:

http://www.intechopen.com/books/trends-in-electromagnetism-from-fundamentals-to-applications/foundations-of-electromagnetism-equivalence-principles-and-cosmic-interactions.

[31] W.-T. Ni, H.-H. Mei and S.-J. Wu, Mod. Phys. Lett. 28 (2013) 1340013.